# The Luminosity-Metallicity Relation for Bulges of Spiral Galaxies

P. Jablonka

*URA 173 CNRS-DAEC, Observatoire de Paris, 92195 Meudon Principal Cedex, France*

P. Martin[1]

*Steward Observatory, University of Arizona, Tucson, Arizona 85721, USA*

and

N. Arimoto

*Institute of Astronomy, University of Tokyo, Mitaka, Tokyo 181, Japan*

## ABSTRACT

Optical integrated spectra of bulges have been obtained for 28 spiral galaxies for which the surface brightness profiles were decomposed into disk and bulge contributions by Kent (1985) and Simien & de Vaucouleurs (1986). By applying an appropriate aperture size for each galaxy, the unavoidable contamination of disk starlight has been carefully minimized and set to $\sim 1/7$ of the total stellar light. The sample covers a wide range in bulge luminosity and morphology. The $Mg_2$ index shows a tight correlation with the bulge intrinsic luminosity, similar - and extended at fainter magnitudes - to the relationship known for ellipticals. Other features such as CaIIK, CN, G-band, and TiO show similar trend with the bulge luminosity. On the contrary, the Fe5270 and Fe5335 iron lines are constant - within some dispersion - for all bulges. A simple application of a stellar population synthesis model suggests that magnesium should be most enhanced with respect to iron in the brightest bulges. Concerning the structural parameters, bulges do occupy the same region in the fundamental plane as ellipticals. In conclusion, the present study favors the idea that the bulk of stars in bulges have formed with only moderate or negligible influence of the disk material, likely by very similar processes as those driving the formation of elliptical galaxies.

*Subject headings:*

---

[1]Current Address: European Southern Observatory, Lasilla 19001, Santiago 19, Chile

## 1. Introduction

Comparisons of properties between galaxies of different Hubble types are important to delineate precisely the key physical processes during galaxy formation.

Bulges of spiral galaxies and elliptical galaxies are generally regarded as stellar systems sharing a number of common dynamical properties. Bulges, like ellipticals, follow the so-called Faber-Jackson relation, even though the quality of the agreement in slope and dispersion depends on whether or not the galaxies have a stellar bar structure (Kormendy & Illingworth 1983). Bulges lie close to the low luminosity ends of the structural parameter relations for bright ellipticals : as ellipticals, more luminous bulges have larger core radii $r_c$, lower surface brightnesses $\mu_0$ and larger central velocity dispersions $\sigma_0$ (Kormendy 1985a,b ; Bender et al. 1992). While the structure of giant ellipticals is mainly supported by the velocity anisotropy, the global dynamics of intermediate ellipticals and bulges is dominated by rotation (Kormendy & Illingworth 1982). Furthermore, isophotal analyses of CCD images of ellipticals have shown that intermediate ellipticals harbor a faint stellar disk (e.g. Bender et al. 1989), drawing a continuous picture from giant ellipticals to spirals, where rotation and disk component become progressively more important.

Surprisingly, stellar populations of bulges have not received much attention. While spectroscopic studies of elliptical galaxies are rather well advanced - ~400 galaxies with central spectroscopic measurements in Faber et al. (1989) - to date, the largest samples of bulges are restricted to ~20 objects (Bica 1988; Bender et al. 1993; Peletier & Balcells 1995). Faber (1973) showed that metal-line strengths of ellipticals increase with galaxy luminosity, and attempts have been made to look for the same behavior in bulges. For instance, Bica (1988) suggested, using population synthesis techniques applied on averaged spectra, that indeed the maximum metallicity of stars in spiral nuclei was related to the bulge luminosity. Bender et al. (1993) compiled data for 25 bulges together with a large sample of elliptical galaxies. However, most of those bulges - except five - had S0 host galaxies. Although it became clear from their work that, in the $Mg_2$ vs $\sigma_0$ plane, bulges of lenticulars show a similar behavior as that of ellipticals, still, it could not be taken as a statement valid for bulges of spirals *in general*. Moreover, for both of these important studies, no real bulge-disk light decomposition has been performed. The consequences of this can hardly be estimated quantitatively, but prevent us from drawing any definitive conclusions on bulge properties.

Indeed, one difficulty in studying bulges of spiral galaxies resides in the fact that even in the galaxy central parts, bulge and disk lights are blended. One can easily check that, in a given aperture, one collects different ratios of bulge and disk light, varying by a large amount from spiral to spiral, due to the large dispersion in galaxy structural parameters (Kent 1985; Simien & de Vaucouleurs 1986). Different distances do also accentuate this effect. No previous work has yet addressed this problem neither handled properly the disk light contamination in the spectroscopic data.

To grasp the nature of stars in bulges and to understand how bulges form, correlations should be sought among properties (luminosity, velocity dispersion, spectral features) of bulge themselves, and not among global properties of spiral galaxies. Also, to evaluate whether disks influence the formation of bulges and to infer how luminosity and chemical enrichment of bulges do relate along the Hubble sequence, spiral galaxies of a wide range in morphologies and intrinsic bulge luminosities should be analyzed. These represent the main goals of the present study.

## 2. Observations and Strategy

Spectral indices of elliptical galaxies were measured in their central regions (Faber et al. 1989; González 1993); comparison purposes led us consequently to observe face-on or moderately inclined spirals. Our spiral sample has been selected from Kent (1985) and Simien & de Vaucouleurs (1986). These authors have decomposed the galactic light into bulge and disk luminosity profiles, fitting an $r^{1/4}$ law to the bulge and an exponential one to the disk. For each galaxy, structural parameters, such as effective radius $r_e$, scale height $h$, surface brightness values $\mu_e$ at $r_e$ and $\mu_0$ at the center were available. Hence, we have been subsequently able to calculate the radius, $r_{L_B/L_D}$, corresponding to a fixed value of the bulge-to-disk luminosity ratio, $L_B/L_D$, taken equal for all galaxies. The explicit forms of $L_B(r)$ and $L_D(r)$, - luminosities within radius r - can be found in the literature (see Mihalas & Binney 1981). Our aim being to fix and minimize the contribution of the disk in our spec-tra, we have defined $r_6$ as the radius within which $L_B/L_D \simeq 6$.

The integrated spectra of 28 spiral galaxies have been obtained during two runs in 1993, using an automatic drift scanning procedure available at the Steward Observatory 2.3m telescope equipped with a Boller & Chivens CCD spectrograph. Each measurement consisted in one integration during which the bulge of the galaxy was scanned back and forth several times. The B&C spectrograph was used with a 12×8k Loral CCD and the 400 gpm grating blazed at 4800 Å to cover the spectral range 3560-6850 Å (a UV-36 blocking filter was also employed). The usable slit length was about 3.5' and three different fixed widths were available: 1.5″, 2.5″ and 4.5″. These apertures selected accordingly with the parameters employed during the scans allowed us to observe precisely the bulge light up to $r_6$. The resulting spectral resolutions correspond to 7.6 Å, 8.6 Å and 11.4 Å, respectively. To minimize spurious effects from atmospheric refraction, the galaxies were observed at very small airmasses (< 1.2). For a few galaxies, no drift scanning was used since $r_6$ was smaller than the larger slit aperture. For these galaxies, the aperture was selected according to $r_6$ and the slit was positioned along the atmospheric dispersion. Integration times were 1800 seconds; when drift scans were necessary, the images crossed the slit at least 8 times. Spectrophotometric standards were observed frequently through the nights using the largest aperture available (4.5″). Following standard procedures, data reduction and analysis were performed using IRAF and the CCDRED and LONG SLIT packages. Flat-fielding using high S/N dome and sky flat-fields was carefully performed and sky subtraction was achieved by defining apertures outside the galaxy light. One-dimensional spectra for the galaxies were then extracted interactively and flux calibrated. Signal-to-noise ratios ∼30-50 have been achieved.

The results of decomposition procedures depend on the photometric band used (de Jong 1995). Kent's (1985) decomposition has been carried out in the r-Gunn band ($\lambda_0$ = 6550 Å and $\Delta\lambda$ = 900 Å ; Thuan & Gunn 1976), while Simien & de Vaucouleurs (1986) have used the Johnson B-band. In this respect, two questions shall be addressed: a) Does $L_B/L_D$ vary along the spectral wavelength range, since bulge and disk may have considerably different spectral energy distribution ? b) How should we relate the $L_B/L_D$ ratios of the 5 galaxies in our sample taken from Simien & de Vaucouleurs data in B-band to the rest of our galaxies, for which $L_B/L_D$ is derived in r-Gunn band ? Taking the Cousin R-band, ($\lambda_0$ = 6400Å and $\Delta\lambda$ = 750Å ) as a fair representative of the r-Gunn band, we have found that while $(L_B/L_D)_B = 6$, $(L_B/L_D)_R = (1.07 \pm 0.3)(L_B/L_D)_B$ and that conversely $(L_B/L_D)_R = 6$, $(L_B/L_D)_B = (0.94\pm0.2)(L_B/L_D)_R$. Thus, errors introduced by using different photometric bands are negligible in our case and $L_B/L_D$ is kept reasonably constant along our wavelength coverage.

In summary, parameters have been chosen so that disk light could not contribute to more than 1/7 of the total integrated light in our spectra. In practice, one had to deal with fixed slit widths, which lead to the final bulge-to-disk ratios displayed in Table 1. Still, the disk-to-total light ratios are confined to 12–18%. Hence, the contamination of disk light has been minimized and all galaxies are analyzed under the same conditions, as much as possible.

In Table 1, we list up general properties of the bulges of our 28 spirals.

## 3. Spectral indice measurements

Because the comparison between bulges and ellipticals was one of our purposes, we have adopted the procedures widely used in recent studies of elliptical galaxies. For the measurement of the equivalent widths of the magnesium and iron lines, we followed the windows and continua defined in González (1993). The other CNO and $\alpha$-elements, such as CaIIK (3930Å), CN (4182Å), G-band (4301Å) and TiO (6242Å) features, were analyzed following the procedures of Bica & Alloin (1986a,b; see also Jablonka et al. 1996). Table 2 gathers the results of the measurements of the molecular and atomic indices. No correction for velocity dispersion is made at this stage. These corrections are only significant for the iron Fe5270 and Fe5335 lines (Gorgas et al. 1990; González 1993).

As the Mg$_2$ indices of our bulges will be directly compared to data obtained for elliptical galaxies in the Lick spectroscopic system (Faber et al. 1989), we compare the indices for a couple of galaxies in common: NGC 2639, NGC 2776, NGC 2916 and NGC 4036 have central indices published in Bender et al. (1993), while Burstein et al. (1988) give Mg$_2$ for NGC 2681. A negligible global difference of $\Delta$ Mg$_2$ = 0.0084 ± 0.0096 is found.

Table 1: General properties of the bulges

| Name | Type | $M_r$ | $r_e$ | $r_6$ | $\sigma_0$ | $L_B/L_D$ |
|---|---|---|---|---|---|---|
| (1) | (2) | (3) | (4) | (5) | (6) | (7) |
| 488 | SAR3 | $-20.88$ | 28.67 | 9.25 | 193 | 6.46 |
| 514 | SXT5 | $-16.30$ | 3.74 | 1.25 | – | 6.87 |
| 524 | LAT+ | $-21.91$ | 47.20 | 19.25 | 242 | 6.00 |
| *628 | SAS5 | $-17.27$ | 25.75 | 5.25 | 65 | 6.62 |
| 670 | .LA0 | $-20.11$ | 15.43 | 5.25 | 99 | 5.69 |
| 697 | SXR5 | $-18.40$ | 15.00 | 1.25 | – | 6.88 |
| 1589 | S..2 | $-20.33$ | 12.35 | 8.25 | – | 6.03 |
| *2268 | SXR4 | $-18.29$ | 5.40 | 4.25 | 76 | 6.26 |
| 2619 | S..4 | $-18.26$ | 10.73 | 0.75 | – | 6.26 |
| 2639 | RSAR1 | $-20.61$ | 28.42 | 2.25 | 188 | 6.01 |
| *2681 | PSXT0 | $-18.62$ | 7.77 | 19.25 | 111 | 6.40 |
| 2683 | SAT3 | $-17.15$ | 71.11 | 3.25 | 140 | 5.71 |
| *2775 | SAR2 | $-20.24$ | 27.95 | 21.25 | 175 | 5.96 |
| 2776 | SXT5 | $-18.38$ | 12.50 | 1.25 | – | 6.27 |
| 2916 | SAT3 | $-17.79$ | 2.11 | 1.25 | – | 8.11 |
| 3053 | SB.1 | $-18.92$ | 5.75 | 4.25 | – | 5.87 |
| 3434 | SAR3 | $-18.80$ | 3.65 | 5.25 | – | 6.04 |
| 3770 | SB.1 | $-18.71$ | 5.29 | 1.25 | – | 7.51 |
| 4036 | .L.. | $-19.25$ | 13.11 | 5.25 | 193 | 6.48 |
| 5440 | S..1 | $-20.43$ | 10.71 | 14.25 | – | 6.01 |
| 5533 | SAT2 | $-21.13$ | 23.19 | 42.25 | – | 5.98 |
| 5676 | SAT4 | $-17.58$ | 4.40 | 2.25 | 112 | 5.11 |
| 5678 | SXT3 | $-17.45$ | 5.85 | 1.25 | – | 4.90 |
| *7217 | RSAR2 | $-19.79$ | 10.84 | 10.25 | 132 | 6.23 |
| 7448 | SAT4 | $-17.95$ | 9.92 | 1.25 | – | 4.83 |
| 7631 | SAR3 | $-18.77$ | 16.75 | 1.25 | 115 | 5.48 |
| 7664 | S..5 | $-18.75$ | 13.73 | 0.75 | 103 | 4.36 |
| 7722 | S..0 | $-21.05$ | 27.42 | 12.25 | – | 6.22 |

Note to Table1:

Columns : (1) galaxy identification : an asterix identifies the galaxies taken from Simien & de Vaucouleurs (1986), the others are taken from Kent (1985); (2) galaxy type from the RC3; (3) bulge total r-band absolute magnitude ; (4) bulge effective radius, in arcsec; (5) scanned radius in arcsec; (6) central velocity dispersion in km s$^{-1}$ from McElroy (1995); (7) bulge-to-disk light ratio within $r_6$.

Table 2: Spectral line indices of the bulges

| Name | Slit | CaIIK | CN | G-band | Mg$_2$ | Fe5270 | Fe5335 | TiO |
|---:|---:|---:|---:|---:|---:|---:|---:|---:|
| NGC | arcsec | Å | Å | Å | mag | Å | Å | Å |
| 488 | 4.5 | 18.43 | 13.82 | 9.80 | 0.30 | 2.57 | 2.29 | 8.66 |
| 514 | 2.5 | 10.18 | 6.73 | 6.57 | 0.17 | 2.50 | 2.17 | 6.12 |
| 524 | 4.5 | 17.14 | 12.67 | 9.47 | 0.30 | 2.48 | 2.30 | 8.98 |
| 628 | 2.5 | 14.19 | 7.60 | 7.61 | 0.18 | 2.72 | 2.59 | 7.42 |
| 670 | 2.5 | 11.45 | 7.86 | 6.56 | 0.22 | 3.10 | 2.77 | 7.28 |
| 697 | 2.5 | 14.00 | 8.08 | 7.67 | 0.19 | 3.33 | 2.93 | 7.27 |
| 1589 | 4.5 | 17.79 | 12.73 | 10.36 | 0.27 | 2.90 | 2.45 | 8.13 |
| 2268 | 4.5 | 14.26 | 7.12 | 7.73 | 0.18 | 2.67 | 2.26 | 7.50 |
| 2619 | 1.5 | 15.20 | 10.84 | 9.65 | 0.19 | 2.78 | 2.53 | 7.65 |
| 2639 | 4.5 | 18.53 | 12.05 | 9.58 | 0.24 | 2.69 | 2.26 | 7.17 |
| 2681 | 4.5 | 12.55 | 5.33 | 6.50 | 0.17 | 2.17 | 2.01 | 4.61 |
| 2683 | 2.5 | 17.66 | 9.10 | 9.58 | 0.23 | 2.84 | 2.68 | 7.59 |
| 2775 | 4.5 | 18.23 | 10.21 | 9.79 | 0.25 | 3.15 | 2.20 | 7.20 |
| 2776 | 2.5 | 13.30 | 7.11 | 7.30 | 0.18 | 2.98 | 2.49 | 7.17 |
| 2916 | 2.5 | 14.47 | 6.46 | 5.07 | 0.22 | 3.10 | 2.56 | 4.80 |
| 3053 | 2.5 | 11.45 | 7.22 | 6.32 | 0.20 | 2.87 | 2.34 | 6.28 |
| 3434 | 4.5 | 6.98 | 5.32 | 5.56 | 0.14 | 2.36 | 1.85 | 5.95 |
| 3770 | 2.5 | 15.52 | 10.37 | 9.43 | 0.23 | 3.30 | 3.03 | 7.83 |
| 4036 | 2.5 | 18.93 | 12.46 | 10.06 | 0.27 | 3.18 | 2.72 | 8.42 |
| 5440 | 4.5 | 17.52 | 10.60 | 9.95 | 0.26 | 2.42 | 2.11 | 9.72 |
| 5533 | 4.5 | 12.51 | 4.61 | 7.42 | 0.18 | 1.93 | 2.02 | 7.73 |
| 5676 | 4.5 | 15.17 | 9.06 | 8.53 | 0.23 | 2.67 | 2.14 | 8.21 |
| 5678 | 2.5 | 10.44 | 4.67 | 6.81 | 0.16 | 2.62 | 1.89 | 6.21 |
| 7217 | 4.5 | 16.12 | 11.12 | 8.70 | 0.28 | 3.04 | 2.45 | 7.41 |
| 7448 | 2.5 | 9.06 | 5.20 | 5.17 | 0.13 | 2.00 | 1.96 | 5.79 |
| 7631 | 2.5 | 13.41 | 6.89 | 7.05 | 0.18 | 1.84 | 2.25 | 7.27 |
| 7664 | 1.5 | 15.86 | 9.20 | 8.72 | 0.20 | 2.58 | 2.13 | 6.39 |
| 7722 | 4.5 | 13.90 | 8.91 | 7.73 | 0.24 | 3.30 | 2.45 | 8.64 |

## 4. The luminosity-metallicity relations

### 4.1. The CNO and $\alpha$-elements

In Fig. 1, we present the variation of the $Mg_2$ index with the bulge absolute magnitude. The bulge magnitudes are taken from Kent (1985) and Simien & de Vaucouleurs (1986). Simien & de Vaucouleurs' B magnitudes were converted into r-Gunn magnitudes by using Kent's (1984) relation. Indeed, magnitudes and $Mg_2$ index are correlated, with a correlation factor of 0.59, which excludes the hypothesis of non-correlation with a 99% confidence. Bulges tend to be fainter in late-type spirals that in early-type ones, but this is the only segregation seen in Fig. 1. Besides this tendency, all Hubble types are mixed. The luminosity is the prime factor of the correlation. The existence of a luminosity-metallicity relation for bulges is now clear : the more luminous bulges are, the more enriched in heavy elements they tend to be.

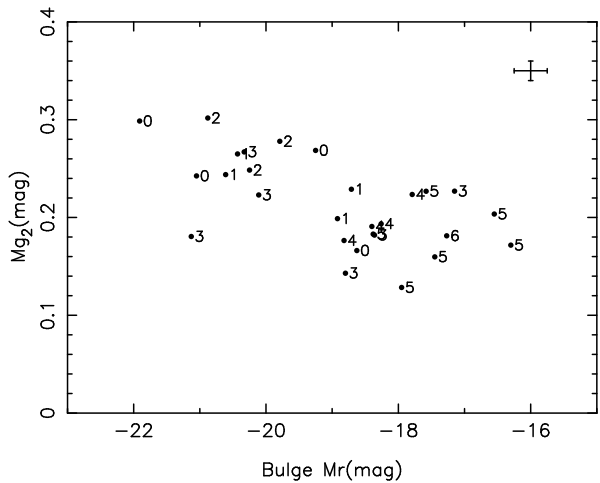

Fig. 1.— The relation between the bulge total r-luminosity and $Mg_2$ index. Numbers represent the galaxies Hubble types (T parameter). Typical maximum errors are shown.

In Fig. 2, we superimpose, to the points of Fig. 1, the mean relation derived for ellipticals by Gorgas & González (1995, private communication) and its $1\sigma$-dispersion in dotted lines, derived after the compilation of Faber et al. (1989) and González (1993) data. This relation has been originally derived in B-band and with a Hubble constant $H_0 = 50$ km s$^{-1}$ Mpc$^{-1}$, while bulge magnitudes were obtained in the r-Gunn band and with $H_0 = 100$ km s$^{-1}$ Mpc$^{-1}$. We have chosen a common value of $H_0 = 100$ km s$^{-1}$ Mpc$^{-1}$, and have transformed the elliptical magnitudes from the B to the r-band, by using Kent (1984) relation: $r = B - 0.10 - 1.18(B-V)$. $B-V$ has been matched by the relation $B - V = 1.12Mg_2 + 0.615$ derived for elliptical galaxies (Burstein et al. 1988). We note that all bulges are within $3\sigma$ of the luminosity - metallicity relation of ellipticals. Slopes and dispersions for bulge and elliptical relations are remarkably identical.

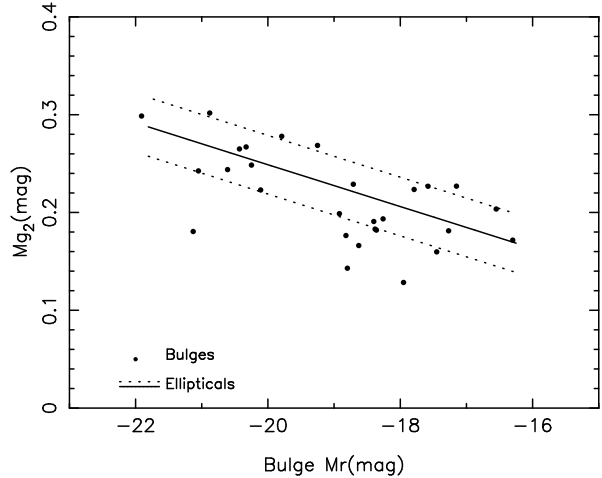

Fig. 2.— The relation between the bulge total r-luminosity and $Mg_2$ index. Plain and doted lines match the mean relation and 1-$\sigma$ dispersion known for ellipticals, respectively.

Figure 3 illustrates our unsuccessful seek for a possible correlation between $Mg_2$ and the disk absolute magnitude. In order to minimize the uncertainties on disk magnitudes via their transformation from B to r-bands, the five galaxies of the Simien & de Vaucouleurs (1986) sample are not included. As discussed by Arimoto & Jablonka (1991), the absolute magnitudes of disks are rather confined to a narrow range around $M_r \simeq -20$ mag with $\Delta M_r \sim 3$ mag, in contrast to the bulge magnitude scatter of $\Delta M_r \sim 6$ mag. Figure 3 clearly shows that no correlation between the disk luminosity and the central $Mg_2$ index is present. Since the disk light contamination has been carefully avoided as much as possible, the bulge $M_r$-$Mg_2$ relation in Figs. 1 and 2 is indeed intrinsic to the stellar population of the bulges themselves and has nothing to do with disk stars.

The $\sigma_0$-$Mg_2$ relation for ellipticals is known to be less dispersed than the luminosity-$Mg_2$ one (Bender et al. 1993). This is partly due to the fact that both quantities are free from uncertainties arising from the distance measurements and partly due to the velocity dispersion reflecting more precisely the gravitational potential than does the stellar luminosity. Unfortunately, $\sigma_0$ values are available only for a sub-sample ($\sim$ half) of our galaxies (McElroy 1995). With our data, we are not in good position to measure the velocity dispersions of the bulges of our sample since by having spatially scanned the galaxies, we would not get the exact central value, and also because of the medium resolution we chose. However, even with this sub-sample, all morphological types of spiral galaxies are represented, and the wide range in bulge luminosity is covered. In Fig. 4, we present the relation between $\sigma_0$ and $Mg_2$ for our bulges. The solid line indicates the relation obtained for elliptical galaxies, as derived by Gorgas & González (1995), and the dashed-lines show its 1$\sigma$-dispersion. Again, bulges and ellipticals have very similar behaviors.

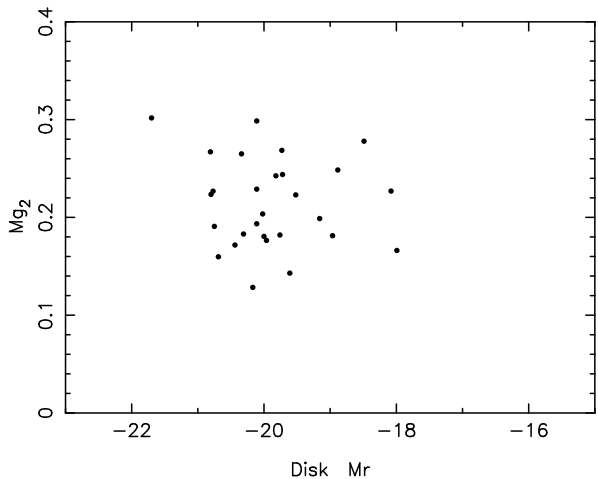

Fig. 3.— The absence of correlation between the disk r-magnitude and the bulge central $Mg_2$ index.

As an example, we have plotted the location of the Galactic bulge in the $\sigma_0$-$Mg_2$ diagram (indicated by the open asterisk). The metallicities of the Galactic bulge stars have been measured by Rich (1988) for 88 K giants in Baade's window. A remarkably wide range of metallicity was revealed. For our purpose, we simply took a median value, [Fe/H] $\sim -0.25$, of the metallicity distribution given by McWilliam & Rich (1994), who recently revised the metallicity calibration of standard stars used by Rich (1988). We then converted this [Fe/H] value into [$\alpha$/H] $\sim 0$ by assuming [$\alpha$/Fe] $\sim 0.3$ (McWilliam & Rich 1994), and obtained $Mg_2 \sim 0.28$, with the help of the formula given by the population synthesis model of Buzzoni et al. (1994). The central velocity dispersion was taken from Rich (1990) as $\sigma_0 = 105 \pm 11$ km s$^{-1}$ for a sample of 53 K giants, in agreement with Sharples et al. (1990) who derived $\sigma = 113 \pm 11$ km s$^{-1}$ for a sample of 225 giant stars. Indeed, the location of the bulge of our Galaxy is consistent with the relationship defined by our sample.

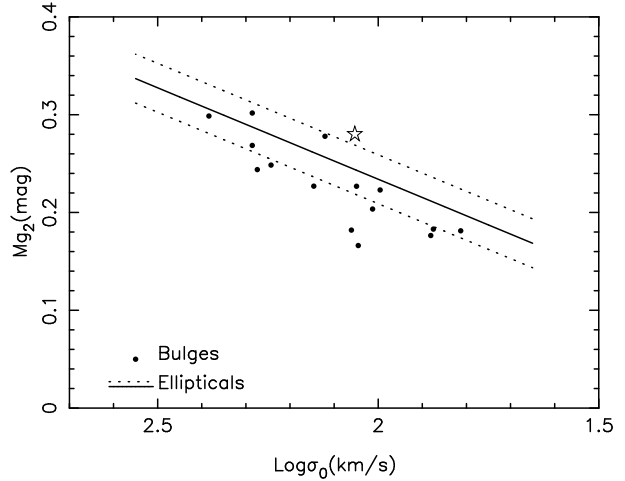

Fig. 4.— The variation of the bulge $Mg_2$ index with the central velocity dispersion $\sigma_0$. Solid and dotted lines are the same as in Fig. 2. Open asterisk indicates the bulge of our Galaxy.

Figure 5 generalizes the luminosity-metallicity relation to CNO and other $\alpha$-elements. CN, CaII K, G-band and TiO show a similar trend with the bulge magnitudes : more luminous bulges are characterized by stronger line indices, likely implying an enhancement of the elements responsible for the formation of these lines.

### 4.2. Iron

The measurement of the equivalent widths of the Fe5270 and Fe5335 lines is sensitive to velocity dispersion and spectral resolution. Our procedure of observation yields *a priori* to non trivial corrections. Elliptical galaxies show radially decreasing profiles of velocity dispersion (e.g., González 1993) and it is very likely that bulges have velocity dispersion profiles of a similar kind. In González (1993) sample, $\sigma_0$ decreases by at most 40 km s$^{-1}$ from the galaxy center to the effective radius. Hence, $\sigma_0$ is indeed a good approximation of the velocity dispersion of the stellar population integrated in our spectra, and are valid for spectral features corrections. As said previously, the published velocity dispersions are available for about a half of our sample. For those galaxies, the iron indices have been accordingly corrected, taking into account the spectral resolution, following Davies et al. (1993).

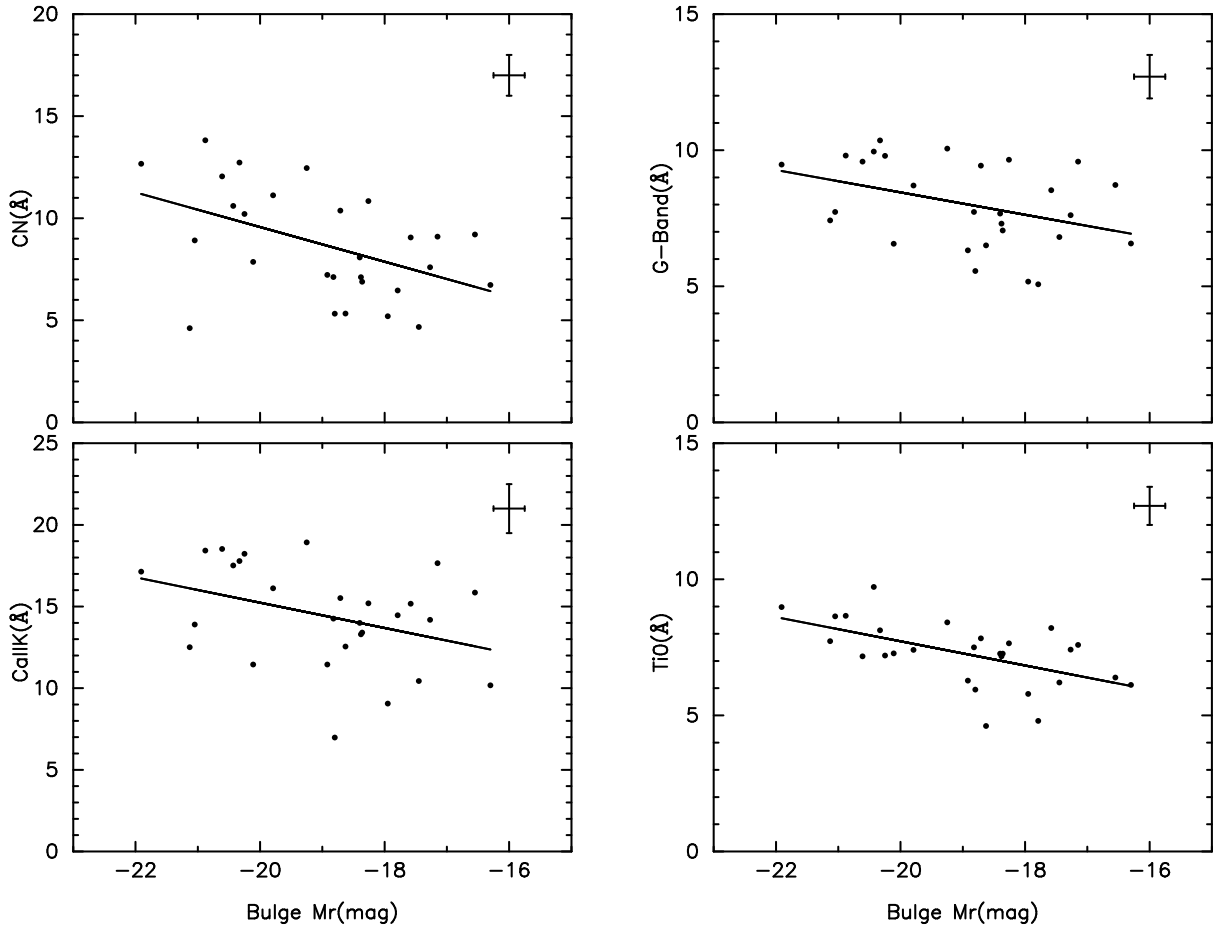

Fig. 5.— Generalization of the luminosity-metallicity relation of bulges for other $\alpha$ and CNO elements. Lines are linear regressions. Typical maximum errors are shown.

Considering the Faber-Jackson relation applied to the other bulges, $\sigma_0$ should reach at most 200km s$^{-1}$ (M$_r$ = $-21$ mag) (Kormendy & Illingworth 1983). As presented in Davies et al. (1993), the correction (multiplying) factors implied by $\sigma_0$= 200km s$^{-1}$ are 1.09 and 1.1 - at 11.4Å and 8.6Å spectral resolution, respectively - for Fe5270 and 1.17 for Fe5335, mostly independently of the spectral resolution; this is the case for four of our galaxies. They are 1.02 for Fe5270 and 1.05 for Fe5335 at 100km s$^{-1}$, independently of the spectral resolution; this is the case for eight galaxies in our sample. We see that such small corrections, if they were applied, would not seriously influence the conclusions given below.

The resulting equivalent widths of Fe5270 are plotted in Fig. 6 as a function of the bulge absolute magnitude. Surprisingly, no correlation emerges. The situation remains the same when Fe5335 is plotted instead. These iron line features are nearly constant and are well confined to a narrow range for all bulges regardless of their luminosities and of the morphological types of their host galaxies, showing a high contrast to other line features that we presented in Figs.2 and 5. This curious behavior has also been pointed out for ellipticals (Faber et al. 1992).

It is not an easy task to assign abundances of ele-ments from analyses of spectral line features, as galaxies are composite stellar systems of velocity dispersions, various ages and metallicities. Furthermore, spectral synthesis methods are premature and still suffer from serious problems in modeling, e.g., uncertainties in the lifetime and luminosities of post main sequence evolutionary stages, the temperature of red giant branch stars, the calibration of color - temperature relation for cool stars, and most seriously the lack of accurate libraries of stellar spectra for cool stars and non-solar metallicities (Charlot et al. 1996). Therefore, one must be cautious when assigning abundances of heavy elements by applying stellar population synthesis models.

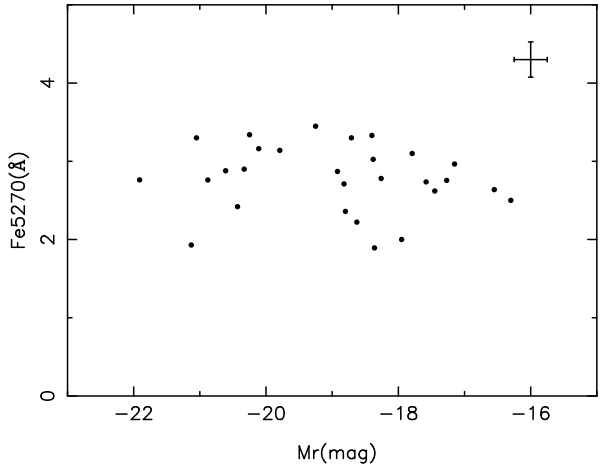

Fig. 6.— The flat dependence of the iron spectral feature with the bulge r-magnitude. Typical maximum errors are shown.

So far, the link between spectral features and abundances had been established by modeling single stellar generations (see Worthey et al. 1992 for a review). This cannot be applied straightforwardly to bulges, as they are composite in metallicity (see Rich 1988 and McWilliam & Rich 1994 for the Galactic bulge). Recently, Barbuy et al. (1995) have synthesized galaxy spectral features by taking into account this distribution of stellar metallicities, following the work of Arimoto & Yoshii (1987). In their model, they assume a supernovae-driven galactic wind, adopt a star formation rate proportional to the gas density and, apply a Salpeter initial stellar mass function. Both Type Ia and II supernovae are explicitly taken into account. A grid of synthetic spectra had been calculated in the range of metallicity $-2.0 < [Fe/H] < +0.5$ with non-solar abundance ratios (Barbuy 1994), which is used for synthesizing galaxy spectra. In Fig. 7, we plot $Mg_2$ and $<Fe>$, an average of Fe5270 and Fe5335 equivalent widths, for our bulge sample and superimpose the prediction calculated from models of Barbuy et al. (1995).

Provided that Barbuy et al.'s predictions are correct, Fig. 7 suggests that bulges are enhanced in magnesium with respect to iron by about a factor $\sim 2$ in average, consistently with the value derived by McWilliam & Rich (1994) for the galactic bulge giants. We display, in Fig. 8, [Mg/Fe] of individual bulges as a function of their luminosities: the more luminous the bulges, the more enhanced in magnesium as compared to iron they tend to be. A simple application of Barbuy et al.'s model gives $<[Fe/H]> \sim 0$ for bulges of any luminosity, while $[Mg/H] \sim 0.1 - 0.2$ for faint bulges ($M_r > -19$ mag) and $[Mg/Fe] \sim 0.5$ for the most luminous bulges ($M_r \simeq -22$ mag). A few attempts have been made to interpret this overabundance in the case of elliptical galaxies, with however no solution showing up clearly (Worthey et al. 1992, Buzzoni et al. 1994, Weiss et al. 1995). So is the situation for bulges as well.

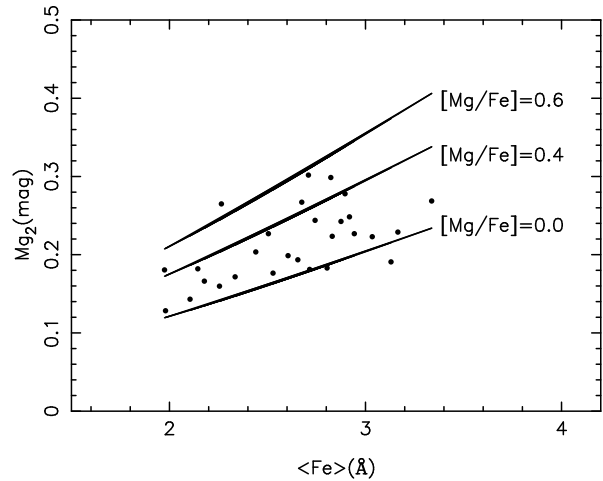

Fig. 7.— The $<Fe>$ vs $Mg_2$ relation for bulges of spirals. The predictions of Barbuy et al. (1995) model are superimposed.

## 5. The Fundamental Plane

While the classical interpretation of the location of galaxies in the fundamental plane (FP) - which deviates from the application of a pure virial equilibrium - invokes the homology of galaxies and the occurrence of dissipation in the course of the galaxy formation (Djorgovski & Santiago 1993), new observations question the hypothesis of homology (Caon et al. 1993; Hjorth & Madsen 1995) and initiate new models of formation without energy dissipation (Capelato et al. 1995).

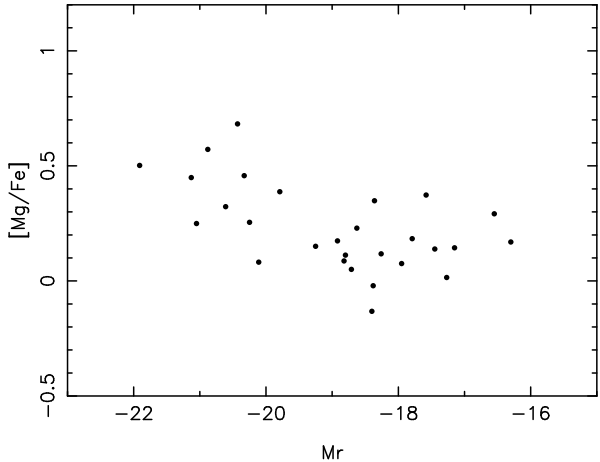

Fig. 8.— The relation between the bulge luminosity and the magnesium over iron abundance.

However, the idea that the FP should tell us about galaxy formation processes is kept. It is therefore of interest to compare the behavior of bulges to that of ellipticals in the FP. Conveniently, Bender et al. (1993) and Simien & de Vaucouleurs (1986) worked both in B-band. This, allowing direct comparison, and the fact that Bender et al. (1993) studied not only ellipticals but also lenticular galaxies in the FP, led us to construct Fig. 9 in the same way as their Fig. 1. In summary, Fig. 9 includes the 17 galaxies of Simien & de Vaucouleurs (1986) sample for which the central velocity dispersions are available. The upper panel of Fig. 9 gives the edge-on view of the FP and the lower one shows it nearly face-on. The precise definitions of global parameters are given in Bender et al. (1992); in short, $\kappa_1 \propto \log M$, $\kappa_3 \propto \log M/L$, and $\kappa_2 \propto \log(M/L)\mu_e^3$. Figure 9 shows that bulges and ellipticals define the same FP and occupy the same domain. On face-on view, bulges fully occupy the area delineated by giant and intermediate ellipticals, down to the bright dwarfs (2 galaxies). So do they in the edge-on view, where they show a comparable dispersion.

Again, bulges seem to form a sequence of objects nearly as large as the family of elliptical galaxies and similarly structured. The physical processes driving up their formation appear quite similar.

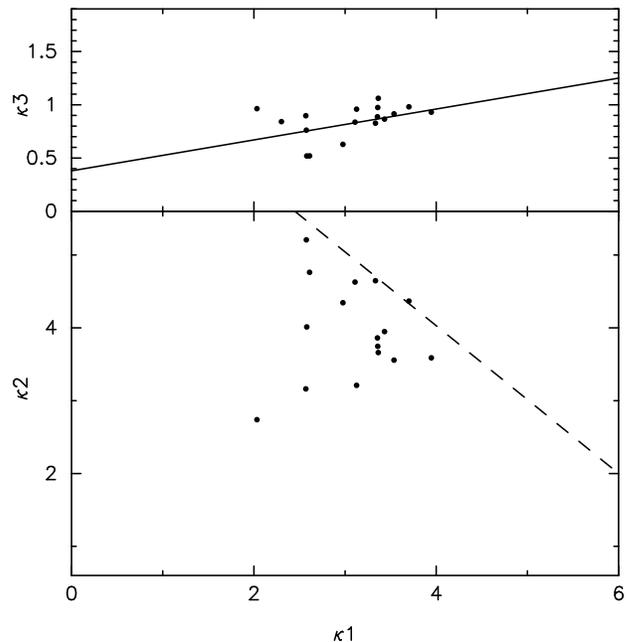

Fig. 9.— The distribution of bulges in the basic global parameter space: central velocity dispersion, surface brightness, and effective radius. The coordinate system ($\kappa_1$, $\kappa_2$, $\kappa_3$) has been chosen to match Fig. 2 of Bender et al. (1992): $\kappa_1 \propto \log M$, $\kappa_2 \propto \log(M/L)\mu_e^3$, $\kappa_3 \propto \log M/L$. (a) Upper panel: the edge-on view of the plane occupied by bulges. (b) Lower panel: nearly face-on view of the plane. The diagonal dashed line highlights the area not occupied by hot stellar systems in Bender et al (1993).

## 6. Discussion and Conclusion

The two main results presented here - a) the luminosity-metallicity relation of bulges similar to that of ellipticals and extending at fainter luminosities, b) the overabundance of magnesium over iron - bring important constraints for models of bulge formation.

On the one hand, a scenario of secular evolution of galaxies along the Hubble sequence by bars has been proposed (see Martinet 1995 for a review). The main idea behind this model is the transformation, via a bar, of a late-type spiral (i.e. with small bulge) into an early-type one with a more prominent spheroidal component. Large amounts of gas driven in the cen-tral region of a galaxy by a stellar bar (e.g. Friedli & Benz 1995) can trigger an intense star formation; newly formed stars in the bulge are then swept out of the galactic plane by vertical instabilities. If enough mass is accreted, the bar itself is destroyed and the resulting galaxy shows a bigger bulge than before bar formation. In this picture, Sd bulges represent proto-bulges. However, as shown in Fig. 8 - if one trusts nowadays synthesis models - [Mg/Fe] is higher for larger luminosities. This is difficult to reconcile with a secular evolution, unless the entire process is completed in less than 1-2 Gyr, time scale for onsets of SNIa explosions (but, recent models by Friedli & Benz (1995) suggest that timescales longer than 1 Gyr seem necessary to transform a small bulge to a bigger one by a bar-driven mechanism.) Indeed, [Mg/Fe] would then tend to decrease instead of the required increase. Besides, the mean value of [Mg/Fe] - $\sim 0.0$ (Pagel 1992) - in the solar neighborhood argues also in favor of rapid decoupling between disk and bulge material. Tsujimoto et al. (1995) in a work with parent assumptions, where bulges are formed from radial inflow of disk gas, have also identified the magnesium overabundance as a clear limitation to their model. We can add that the most luminous (massive) bulges in our sample are clearly built up by old stellar population with no trace of intermediate age stars. This discards a continuous star formation over a long period. Finally, it is non-trivial to understand how different formation processes between ellipticals and spirals would lead to the same relation between mass and chemical enrichment.

On the other hand, Ortolani et al. (1995) have shown that bulge globular clusters and Baade's Window stars of our Galaxy are likely as old as halo globular clusters, supporting the idea of rapid formation and chemical enrichment of the bulge, early in the evolution of the Galaxy. The present study favors such an interpretation : the bulk of star formation in bulges happens likely *before* disk material and bars influence their evolution.

We wish to thank Drs. B. Barbuy and R.S. de Jong for fruitful discussions. We are grateful to Drs J.J. González and J. Gorgas for letting us using their data prior to publication. This work was financially supported in part by a Grant-in-Aid for the Scientific Research (No.06640349 and No.07222206) by the Japanese Ministry of Education, Culture, Sport and Science. Pierre Martin was supported by a NSERC (Canada) Postdoctoral Fellowship, and in part by the NSF through grant AST 90-19150.

12